\documentclass[fleqn,10pt]{wlscirep}
\usepackage[utf8]{inputenc}
\usepackage[T1]{fontenc}
\usepackage{amssymb}
\usepackage{siunitx}
\usepackage{multirow} 

\title{Corneal Pachymetry by AS-OCT after Descemet's Membrane Endothelial Keratoplasty}

\author[1,*]{Friso G. Heslinga}
\author[1]{Ruben T. Lucassen}
\author[1]{Myrthe A. van den Berg}
\author[1]{Luuk van der Hoek}
\author[1]{Josien P.W. Pluim}
\author[2,3]{Javier Cabrerizo}
\author[2]{Mark Alberti}
\author[1]{Mitko Veta}

\affil[1]{Department of Biomedical Engineering, Eindhoven University of Technology, Eindhoven, the Netherlands}
\affil[2]{Ophthalmology Department, Rigshospitalet - Glostrup, Copenhagen, Denmark}
\affil[3]{Copenhagen Eye Foundation, Copenhagen, Denmark}

\affil[*]{fgheslinga@gmail.com}


\begin{abstract}
Corneal thickness (pachymetry) maps can be used to monitor restoration of corneal endothelial function, for example after Descemet’s membrane endothelial keratoplasty (DMEK). Automated delineation of the corneal interfaces in anterior segment optical coherence tomography (AS-OCT) can be challenging for corneas that are irregularly shaped due to pathology, or as a consequence of surgery, leading to incorrect thickness measurements. In this research, deep learning is used to automatically delineate the corneal interfaces and measure corneal thickness with high accuracy in post-DMEK AS-OCT B-scans. Three different deep learning strategies were developed based on 960 B-scans from 50 patients. On an independent test set of 320 B-scans, corneal thickness could be measured with an error of 13.98 to 15.50 $\mu$m for the central 9~mm range, which is less than 3\% of the average corneal thickness. The accurate thickness measurements were used to construct detailed pachymetry maps. Moreover, follow-up scans could be registered based on anatomical landmarks to obtain differential pachymetry maps. These maps may enable a more comprehensive understanding of the restoration of the endothelial function after DMEK, where thickness often varies throughout different regions of the cornea, and subsequently contribute to a standardized postoperative regime.
\end{abstract}
\begin{document}

\flushbottom
\maketitle

\thispagestyle{empty}

\section*{Introduction}
\begin{figure}[t]
\centering
{\includegraphics[width=\linewidth]{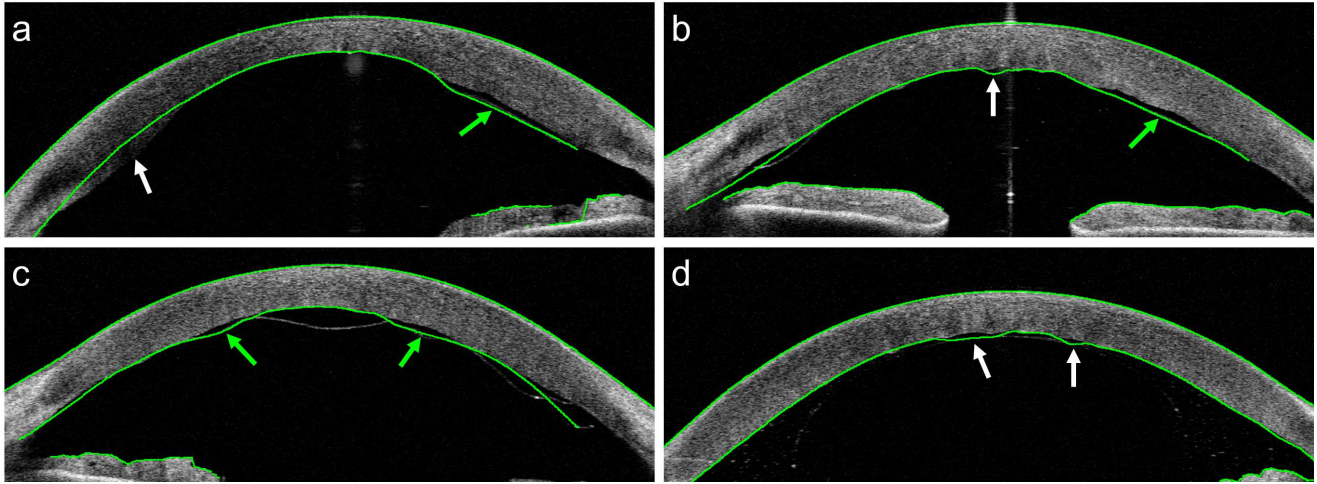}}
\caption{AS-OCT B-scans collected from patients after DMEK surgery. The green lines represent delineations of the (corneal) interfaces by the built-in software of the OCT system. These examples were selected to show the types of delineation errors encountered. In (a), (b), and (c) the delineation partly follows the DMEK graft (green arrows) instead of the posterior interface. Other types of mistakes are indicated by white arrows: (a) Some of the posterior part of the cornea is missed. (b) The delineation does not follow the irregularly shaped interface in the center. (d) The system confuses the boundaries of the gas bubble used in DMEK with the posterior corneal interface.}
\label{fig:casia_examples}
\end{figure}

Corneal thickness is a key biomarker for corneal disorders, including Fuchs’ endothelial dystrophy \cite{Kopplin2012,Patel2020}, keratoconus \cite{Ambrosia2006,Li2008}, and keratitis \cite{Cook1953,Wilhelmus2006}. Measurements on the corneal thickness, called pachymetry, enable detection of thickness changes that are indicative of restoration of corneal endothelial function after surgical treatment. For visualization of the restoring cornea, anterior segment optical coherence tomography (AS-OCT) has become the preferred imaging modality due to its high resolution and reproducibility \cite{Lim2015,Wang2019}. While current OCT software works well for delineating the boundaries of healthy corneas, it often fails for
corneas that are irregularly shaped due to pathology, or as a consequence of surgery (Figure \ref{fig:casia_examples}). Manual correction of the delineation mistakes is time consuming and not practical for a clinical setting. 

In recent years, automated image analysis using deep learning \cite{Lecun2015} has shown to be promising for ophthalmic applications \cite{Ting2019}, including the analysis of AS-OCT images \cite{Xu2020,Fu2019,Treder2019,Heslinga2020}. Deep learning is a subset of machine learning techniques with models that contain many (typically millions of) trainable weights. These weights are iteratively updated with respect to some loss function which compares model predictions to ground truth labels. In contrast with classical machine learning techniques, no handcrafted features have to be selected as the relevant features are automatically learned from the (image) data. Recent work already showed the potential of deep learning for corneal pachymetry by AS-OCT, specifically for keratoconus\cite{Santos2019}. We hypothesized that a similar approach could be used for corneal pachymetry for cases with irregular inner corneal curvature and/or structures that look similar to the corneal boundaries, both of which can lead to delineation failures by standard AS-OCT software.

In this study we focus on OCT scans acquired after Descemet’s Membrane Endothelial Keratoplasty (DMEK) \cite{Melles2006}. During DMEK, the diseased corneal endothelium and Descemet's membrane are replaced with a donor graft. After placement of the graft, a gas bubble is injected into the anterior chamber to support graft attachment to the host cornea. Both the procedural gas bubble and donor graft can mimic the appearance of the corneal interface and result in incorrect delineation.

\begin{figure}[t!]
\centering
{\includegraphics[width=15cm]{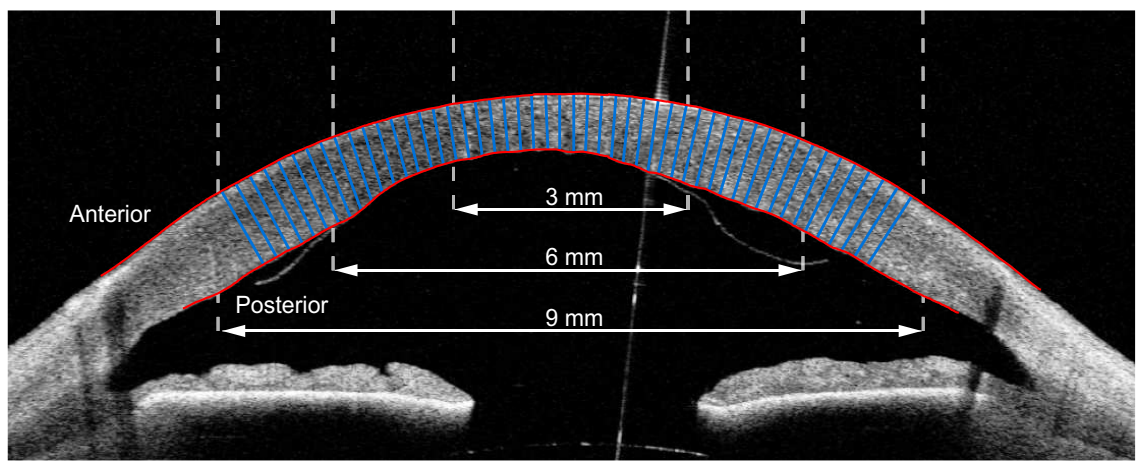}}
\caption{Single image (B-scan) from an AS-OCT scan, showing the cornea and the anterior chamber. This B-scan was cropped centrally and horizontally aligned as reported by Heslinga \& Alberti \cite{Heslinga2020}. Manual delineations of the corneal interfaces are shown in red. Corneal thickness is measured as the distance between the anterior and posterior interface, perpendicular to the anterior interface. The blue lines illustrate a subset of these thickness measurements. For evaluation of the thickness measurements, we distinguish the central 3~mm, 6~mm, and 9~mm diameter with respect to the corneal apex.}
\label{fig:thickness}
\end{figure}

We compare three different deep learning techniques that were developed or used for ophthalmology applications and shown to be highly effective. We validate our thickness measurements for the central 9~mm diameter (Figure \ref{fig:thickness}), whereas previous work only did so for 3.1~mm \cite{Santos2019}. This is essential to assess corneal regeneration after DMEK surgery which uses a graft of $\sim$8.5~mm. In addition, we present an automatic approach for reconstructing differential pachymetry maps that locates the center of the cornea in subsequent images and visualizes thickness differences over time.

\section*{Results}
\subsection*{AS-OCT data \& annotations}
The AS-OCT scans used in this study were collected as part of a randomized controlled trial conducted at the Department of Ophthalmology, Rigshospitalet – Glostrup, Denmark. The trial was designed to compare air and sulfur hexafluoride (SF6) DMEK surgery in patients with Fuchs' endothelial dystrophy or pseudophakic bullous keratopathy \cite{Alberti_RCT}. Repeat DMEK procedures and patients with prior keratoplasty were excluded. A total of 80 swept-source AS-OCT scans \emph{(CASIA2; Tomey Corp. Nagoya, Japan)} from 68 participants were acquired either immediately after surgery, one week after surgery, or both. Each scan consists of 16 images (B-scans) acquired in a radial pattern, corresponding to 1280 B-scans in total.

AS-OCT scans were preprocessed similar as reported by Heslinga \& Alberti \cite{Heslinga2020}. In brief, a deep learning-based localization model was applied to each B-scan to identify the scleral spur, a landmark in the anterior chamber of the eye \cite{Ang2018}. Per full AS-OCT scan, an ellipse was fitted through the scleral spur points of all 16 B-scans to ensure that the points lie in the same plane and to refine point locations. B-scans were horizontally aligned and cropped based on the scleral spur locations, centering around the corneal apex (Figure \ref{fig:thickness}). Final crop sizes were 960~by~384 pixels (width by height) with a pixel size of \SI{15.0}{\micro\metre}. For a detailed description of the participant characteristics and processing methodology, we refer the reader to \cite{Heslinga2020}.

For each B-scan, the anterior corneal interface was annotated inside a 12~mm diameter from the radial center. A diameter of 10~mm was used for the posterior interface. Partial DMEK graft detachments were excluded from posterior interface annotations. The data set was randomly split on a participant level in a training set of 752 images, a validation set of 208 images and a test set of 320 images. B-scans of the training and validation set were annotated by one of three observers under supervision of a cornea specialist. The test set was annotated by all three observers to assess inter-observer variability.

\subsection*{Thickness measurements}
Three deep learning-based models were trained to locate the anterior and posterior corneal boundaries: (1) a patch-based convolutional neural network (CNN), (2) a U-Net\cite{Ronneberger2015} based model, and (3) a CNN with dimension reduction. Details about the model architectures and training process are provided in the methods section. 

Corneal thickness was measured perpendicularly to the anterior interface (see Figure \ref{fig:thickness}), similar to \cite{Li2006}. For each B-scan of the test set, we evaluated thickness for every pixel on the anterior interface inside a 3~mm, 6~mm, and 9~mm diameter. Corneal thickness estimates by the deep learning models were compared with all three sets of annotations (960 in total). Mean absolute errors (MAE) (shown in Table \ref{Tab:MAD-distance}) are very similar for the three deep learning models and across different diameters. The smallest error was found for the U-Net model for the 6 mm diameter (13.84 \SI{}{\micro\metre}), while the largest error was found for the patch-based CNN for the 9~mm diameter (15.50 \SI{}{\micro\metre}). Apart from the latter, all mean absolute errors are smaller than one pixel (15.0 \SI{}{\micro\metre}). The small standard deviations shown in Table \ref{Tab:MAD-distance} show the high repeatability over multiple training runs. In addition, we calculated the standard deviation over the mean absolute errors across the B-scans. Averaged over five training runs, these standard deviations for the central 9~mm diameter were 4.35 \SI{}{\micro\metre} (patch-based), 4.77 \SI{}{\micro\metre} (U-Net), and 4.90 \SI{}{\micro\metre} (CNN with dimension reduction).

\begin{table}[t!]
    \centering
    \caption{Mean absolute error in \SI{}{\micro\metre} of corneal thickness predictions on test set. Comparisons represent deep learning models vs. annotations (left) and annotator vs. annotator (right). Mean~$\pm$~SD of 5 training repetitions.}
    \begin{tabular}{lcccccc}
        \toprule
        \multirow[t]{2}{*} &
        \multicolumn{3}{c}{Models vs. annotations} & \multicolumn{3}{c}{Inter-observer comparison}\\
        \cmidrule(r){2-7}
        Diameter & Patch-based CNN & U-Net & CNN with dim. red. & 1 vs 2 & 1 vs 3 & 2 vs 3\\
        \cmidrule(r){2-4} \cmidrule(l){5-7}
        3 mm& 14.40~$\pm$~0.69 & 13.94~$\pm$~0.38 & 13.94~$\pm$~0.25 & 23.49 & 14.69 & 17.95 \\
        6 mm& 14.80~$\pm$~0.51 & 13.84~$\pm$~0.22 & 14.17~$\pm$~0.22 & 23.71 & 13.91 & 18.80 \\ 
        9 mm& 15.50~$\pm$~0.59 & 13.98~$\pm$~0.15 & 14.40~$\pm$~0.15 & 23.39 & 13.66 & 19.26 \\
        \bottomrule
    \end{tabular}
    \label{Tab:MAD-distance}
\end{table}

We investigated inter-observer variability by comparing the corneal thickness annotations between observers. The results of this comparison are shown in Table \ref{Tab:MAD-distance}. Only for the combination of observer 1 vs 3, the mean absolute error is similar to that obtained with the deep learning models (13.66 - 14.69 \SI{}{\micro\metre}), while the differences between the other combination of observers are substantially larger (17.95 - 23.71 \SI{}{\micro\metre}).
In addition, a cornea specialist manually assessed all delineations of the test set by the built-in software (version 3H.1) of the OCT system. In 90 out of 320 (28\%) B-scans a delineation mistake occurred that resulted in a clinically relevant thickness inaccuracy within the central 9 mm. Out of these 90 B-scans, 74 (82\%) included an inaccuracy that overlapped with the location of a partial DMEK graft detachment. For 37 (12\%) B-scans the thickness errors were considered severe. Sixteen out of 20 (80\%) AS-OCT scans contained at least one B-scan with a clinically relevant thickness inaccuracy, resulting in an incorrect pachymetry map.

\subsection*{Outlier analysis}
\begin{figure}[ht]
\centering
{\includegraphics[width=\linewidth]{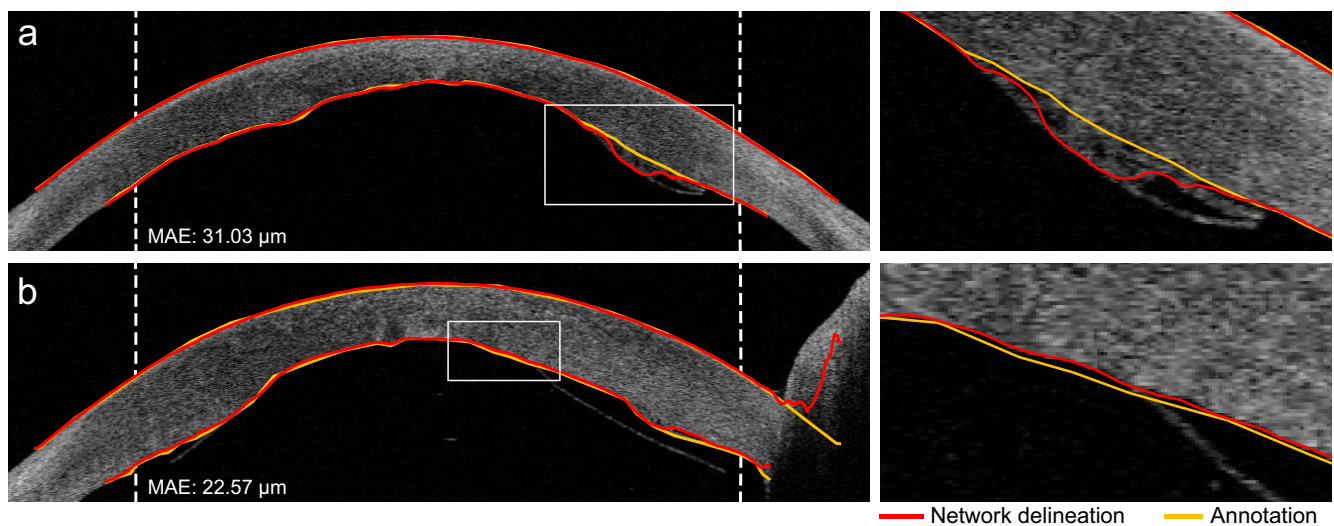}}
\caption{Two examples of B-scans including the annotations and delineations by the CNN with dimension reduction with substantial deviations in predicted thickness. The rectangular areas are enlarged and displayed to the right of the B-scan. Vertical dashed lines indicate the 9~mm diameter. Note that the thickness was not evaluated outside of the 9~mm diameter.}
\label{fig:examples}
\end{figure}

We further inspected the origin of the deviations in thickness measurements between the deep learning models and the manual annotations by investigating the cases with the largest deviations. Figure \ref{fig:examples} shows two example outliers with annotations and delineations by the CNN with dimension reduction.
In Figure \ref{fig:examples}a the remnant tissue from the host or donor prevents the graft from completely attaching at the right posterior side of the cornea. The tissue was correctly excluded from the annotation, but included in the delineation by the network. 

Another example is presented in Figure \ref{fig:examples}b, where the enlarged region shows a shortfall in detail of the annotation compared to the network delineation. Note that the graft is not entirely attached at the right side of the posterior interface, which was correctly recognized by the network and mistakenly included in the annotation.

\subsection*{Pachymetry mapping}
\begin{figure}[h]
\centering
{\includegraphics[width=\linewidth]{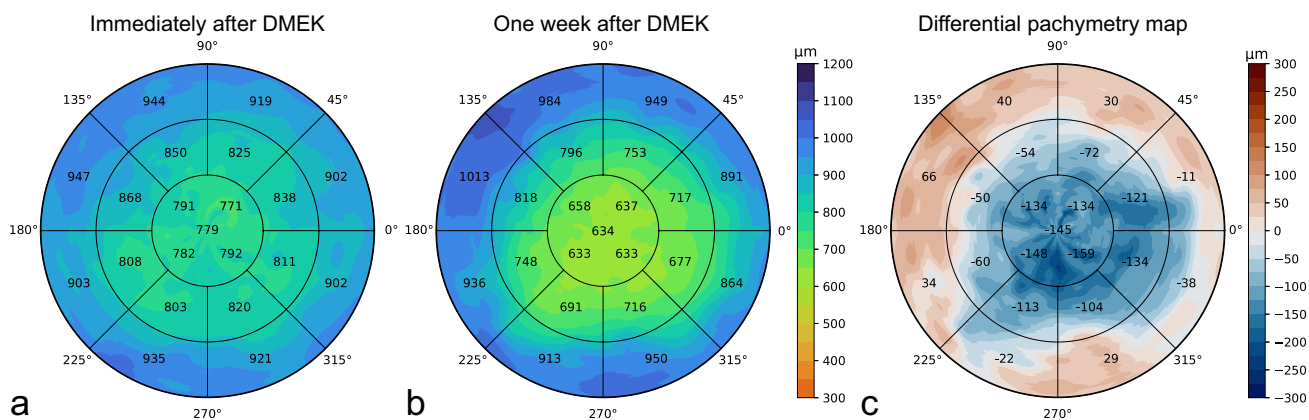}}
\caption{Example of pachymetry maps from one participant in the test set. (a) Pachymetry map of AS-OCT scan acquired immediately after DMEK; (b) Pachymetry map of AS-OCT scan acquired one week after DMEK; (c) Differential pachymetry map of difference in corneal thickness between (b) and (a).}
\label{fig:maps}
\end{figure}

Corneal thickness measurements from 16 radial B-scans were combined to construct pachymetry maps as shown in Figure \ref{fig:maps}. Thickness measurements were plotted on a polar coordinate axis with cubic interpolation between the cross-sections. The pachymetry map was divided into three circular regions with diameters of 3~mm, 6~mm, and 9~mm. The outer two rings were divided into octants where the average thickness is displayed. The inner circle displays the average of the four quarters, as well as the average apex thickness inside the central 1~mm diameter. Thickness values were mapped to corresponding colors of a discrete colormap. Similar to conventional pachymetry colormaps the corneal thickness at 600 \SI{}{\micro\metre} is displayed in green, thinner regions in red, and thicker regions in blue \cite{Bourges2009,Patel2020}.

\subsection*{Effect of training set size}
The precise manual annotation of the corneal boundaries is a time-consuming process. For future projects it is useful to know whether a smaller set of annotated data can be used to obtain similar results. We therefore investigated the effect of the size of the annotated training set on the quality of the thickness measurements. Table \ref{Tab:Training_size} shows thickness measurement errors on the test set B-scans when the deep learning models are trained with 100\%, 50\%, 25\%, and 10\% of the original training set. For the patch-based CNN and the CNN with dimension reduction we found only a marginal increase in the mean absolute error (of about 2 \SI{}{\micro\metre}) for the whole central 9~mm range, when trained with only 10\% of the data. In contrast, the U-Net model performance did decrease more substantially for the 9~mm range when trained with 10-50\% of the data. Further inspection learned this was caused by some substantial error in the 6-9~mm range for a small portion of the B-scans.

\begin{table}[ht]
    \centering
    \caption{Mean absolute error in \SI{}{\micro\metre} of corneal thickness predictions on test set for varying partitions of the training set. Mean~$\pm$~standard deviation of 5 training repetitions.}
    \begin{tabular}{lccccccccc}
        \toprule
        \multirow[t]{2}{*} &
        \multicolumn{3}{c}{Patch-based CNN} & \multicolumn{3}{c}{U-Net} & \multicolumn{3}{c}{CNN with dimension reduction}  \\
        \cmidrule(r){2-10}
        Diameter &  3 mm   &  6 mm   &  9 mm &  3 mm   &  6 mm   &  9 mm &  3 mm   &  6 mm   &  9 mm\\
        \cmidrule(r){2-4} \cmidrule(lr){5-7} \cmidrule(l){8-10}
        100\% & 14.40       & 14.80       & 15.50       & 13.94       & 13.84       & 13.98       & 13.94       & 14.17       & 14.40 \\
              & ~$\pm$~0.69 & ~$\pm$~0.51 & ~$\pm$~0.59 & ~$\pm$~0.38 & ~$\pm$~0.22 & ~$\pm$~0.15 & ~$\pm$~0.25 & ~$\pm$~0.22 & ~$\pm$~0.15 \\
        50\%  & 14.28       & 14.65       & 15.23       & 14.36       & 14.30       & 17.25       & 15.04       & 15.10       & 15.24 \\
              & ~$\pm$~0.33 & ~$\pm$~0.39 & ~$\pm$~0.43 & ~$\pm$~0.51 & ~$\pm$~0.46 & ~$\pm$~5.97 & ~$\pm$~0.69 & ~$\pm$~0.47 & ~$\pm$~0.36 \\
        25\%  & 15.04       & 15.16       & 15.71       & 16.22       & 15.25       & 20.53       & 15.03       & 14.92       & 15.12 \\
              & ~$\pm$~1.13 & ~$\pm$~0.95 & ~$\pm$~0.89 & ~$\pm$~1.45 & ~$\pm$~0.86 & ~$\pm$~7.94 & ~$\pm$~0.49 & ~$\pm$~0.25 & ~$\pm$~0.20 \\
        10\%  & 15.93       & 16.08       & 16.76       & 16.19       & 15.35       & 21.12       & 16.76       & 15.88       & 16.08 \\
              & ~$\pm$~1.30 & ~$\pm$~1.39 & ~$\pm$~1.73 & ~$\pm$~1.38 & ~$\pm$~0.74 & ~$\pm$~6.24 & ~$\pm$~1.18 & ~$\pm$~0.87 & ~$\pm$~0.76 \\
        
        \bottomrule
    \end{tabular}
    \label{Tab:Training_size}
\end{table}


\section*{Discussion}

This research shows the feasibility of automated corneal thickness measurements in post-DMEK AS-OCT scans using deep learning. While our data set contains many examples of irregularly shaped corneas and partly detached DMEK grafts, all models are able to measure corneal thickness with an average error of 13.98 to 15.50 \SI{}{\micro\metre} for the central 9~mm range. In comparison with a typical central corneal thickness of 540 \SI{}{\micro\metre} \cite{Ma2018,Hashemi2011}, this corresponds to an error of less than 3\%. The quality of our thickness measurements is at least on par with manual annotations, as indicated by the inter-observer distance of 13.66 to 23.39 \SI{}{\micro\metre} for the central 9~mm range. 
Based on the accurate thickness measurements, detailed pachymetry maps can be constructed. The preprocessing based on the scleral spur locations largely eliminates spatial translation in the coronal plane and sagittal rotation between follow-up scans\cite{Heslinga2020}. Our method does not correct for coronal rotation, but we expect the effect of this type of head tilt or eye rotation to be small. 

Since follow-up scans can be registered, differential pachymetry maps can be constructed to monitor thickness changes. This may enable a more comprehensive understanding of the restoration of the endothelial function after DMEK, where thickness often varies throughout different regions of the cornea and the restoration of corneal thickness is associated with success of the procedure\cite{Melles2020}. Typically, only the central corneal thickness (CCT) is reported, while this single parameter does not necessarily reflect restoration of the full cornea after DMEK. The DMEK graft is about 8.5 mm \cite{Rock2015} and partial graft detachment happens most frequently in the peripheral region \cite{Bucher2015,Deng2015}. Detachment can sometimes be ambiguous, even in high-quality OCT imaging, as the graft can be close to the inner cornea yet not attached\cite{Heslinga2020}. Local changes in corneal thickness could then be indicative of corneal restoration and thus graft attachment. The differential pachymetry maps presented in this research enable both qualitative and quantitative progression tracking options within the central 9~mm range, covering the whole region of the DMEK graft.

The data for this study only included post-DMEK AS-OCT scans of patients with previous Fuchs’ endothelial dystrophy or pseudophakic bullous keratopathy. The application of the here presented deep learning models for scans from other pathologies or taken in different centra requires further research. 

Three deep learning methods were compared in this research. Despite substantial differences in the approaches, the results for the different models were all at least on par with manual annotations. This could indicate that delineating the corneal boundaries is well-defined and relatively easily solvable using deep learning. This finding is in line with other research aiming at delineating layers in ophthalmic OCT imaging \cite{Santos2019,Wang2020,Liefers2019}. Based on the thickness results for the models trained with 100\% of the training data, none of the models seems to outperform the other. However, construction of a pachymetry map with the patch-based CNN takes considerably longer because of the large number of patches involved and the extensive post-processing steps.

Results of the deep learning models could not directly be compared with the delineations by the built-in software of the OCT system. Nevertheless, 28\% of B-scans were found to contain delineations mistakes by the built-in software leading to a clinically relevant thickness inaccuracy. Moreover, 80\% of the AS-OCT scans of the test set contained at least one B-scan with a thickness inaccuracy of clinical relevance. We observed that these errors often occurred at locations of high clinical relevance, such as an irregularly shaped corneal center, or where the DMEK graft detached.

Training the models with smaller partitions of the data provides insight in the value of adding extra annotated data. Since no improvements were obtained by using 100\% of the available training data compared to only 50\%, it can be concluded that the performance has saturated, and additional data would not contribute to much further improvement. An exception could be the addition of data from rare cases or examples that led to errors in the current test set (e.g. Figure \ref{fig:examples}a). For the U-Net model, training with 10\% or 25\% of the training data did result in an increase in the thickness error in the 6 to 9~mm region. Further inspection revealed that this was due to a small number of B-scans for which the segmentation failed. Interestingly, for the CNN with dimension reduction and the patch-based CNN, even training with 80 B-scans from 5 patients does not seem to reduce the performance of the thickness measurements substantially (MAE of 16.08 \SI{}{\micro\metre} and 16.76 \SI{}{\micro\metre} respectively). These results indicate that future projects on delineation of corneal boundaries could already be developed with less annotated data, yet still obtain reasonable results.

The high resolution of the AS-OCT combined with deep learning for automated image processing supports fast and accurate analysis of the corneal thickness after DMEK. The here presented (differential) pachymetry maps enable tracking of local corneal thickness changes indicative of corneal restoration. As such, these tools can contribute to the ongoing research efforts towards further improvement of the DMEK procedure and management of the postoperative regime. 


\section*{Methods}

\subsection*{Models \& Training}
We implemented three different deep learning models based on recent successful applications related to segmentation or interface delineation in ophthalmic OCT imaging. The first model is based on the patch-based approach by Wang \textit{et al}. for the identification of five different retinal interfaces in spectral domain OCT images\cite{Wang2020}. Using patches of $33\times33$ pixels and a relatively shallow network of 5 trainable layers, Want et al. achieved localization accuracies of 89-98\%. The second model is based on the U-Net architecture which has become the de facto standard method for medical image segmentation \cite{Ronneberger2015}. Multiple adaptations of U-Net were evaluated by dos Santos et al. to segment three corneal layers in images captured with a custom-built ultrahigh-resolution OCT system\cite{Santos2019}. Using cross-validation, a mean segmentation-accuracy of 99.56\% was achieved. The third model was also inspired by U-Net, but modified by Liefers et al. to reduce the dimensionality and output one-dimensional arrays with $y$-locations of three retinal layers in OCT images\cite{Liefers2019}. For the localization of these retinal layers, the authors obtained a mean absolute difference between the predictions and annotations of 1.31 pixels. 

For our application of corneal interface localization, both the patch-based approach and the CNN with dimension reduction allow for direct delineation of the corneal interfaces. In contrast, a U-Net approach is used to segment the cornea and requires a post-processing step to obtain the interface delineations from the segmented mask.
Details about the model adaptations, implementations, and training procedures are described below for each model. Depending on the model requirements we also adapted the preprocessing and post-processing steps. All models were implemented in Keras\cite{chollet2015} with TensorFlow\cite{tensorflow2015} backend and optimized using Adam\cite{Kingma2014}. 

\subsubsection*{Patch-based CNN}
The architecture of the patch-based model was similar to that of Wang et al.\cite{Wang2020} with 2 modifications: (1) $5\times5$ convolutional layers were replaced by two $3\times3$ convolutional layers as factorization is considered more efficient \cite{Szegedy2016}; (2) $3\times3$ average pooling operations in the final two layers were replaced with $2\times2$ max pooling operations. From the full images of training and validation sets, patches of $33\times33$ pixels were extracted for each $x$-coordinate where the interfaces had been annotated (center 12~mm for the anterior interface and center 10~mm for the posterior interface). Anterior and posterior patches were sampled using the respective annotations as center pixel locations. Similarly, for each $x$-coordinate within the central 12~mm, a non-interface patch was constructed for one random pixel not part of the interface annotations. All patches (1.70 million for the training set and 0.47 million for the validation set) were extracted prior to training to speed up the training process.

The model was optimized by minimizing the categorical cross-entropy between the pixel ground truth and model predictions. Online data augmentation was added by rotating the patches with a maximum of 30 degrees. Based on preliminary experiments, the model was trained for 20 epochs with a variable learning rate: 0.001 for epoch 1 to 12, 0.0001 for epoch 13 to 16, and 0.00001 for epoch 17 to 20.

For evaluations on the test set, patches were extracted for all pixels within the center 12~mm (width) and processed by the trained model, predicting either anterior interface, posterior interface, or background for the center pixel of the patch. Based on preliminary results on the validation set, the following post-processing steps were performed for the pixels identified as interface: (1) small connected regions (0 - 250 pixels) were removed; (2) the largest connected region was considered to be true; (3) other regions were considered to be part of the true prediction only when those would be at the same height as the largest connected region; (4) per interface, $y$-values of positive pixels were averaged to obtain a single value per $x$-coordinate; (5) any remaining gaps were filled using linear interpolation of adjacent interface locations.

\subsubsection*{U-Net}
As an alternative to directly delineating the interfaces, a U-Net\cite{Ronneberger2015} was implemented to segment the whole cornea. The U-Net consisted of the standard 4 downsampling (and upsampling) segments and we included batch normalization and residual layers to accelerate training. Binary masks of the cornea were created using the interface annotations. As a preprocessing step, the original images and masks were cropped to 800~by~256 pixels (width by height) and split into a superior and inferior half. This step was included to reduce the size of the input to the U-Net while doubling the number of training examples.

For optimization of the U-Net we experimented with different loss functions (Dice, binary cross-entropy, and weighted binary cross-entropy) and learning rate schedules. We found only minor differences in the results on the validation set and used binary cross-entropy for the final model. We trained for 30 epochs with an initial learning rate of 0.001 that was divided by two at every 3 epochs. We also experimented with data augmentation (brightness adaptation and addition of Gaussian noise) but did not identify any improvements on the validation results.

For evaluations on the test set, the inferior an posterior crops were processed by the trained U-Net and combined. From the predicted mask, the maximum and minimum $y$-values were used to reconstruct the anterior and posterior interface respectively. In contrast to the patch-based model, the predicted $y$-values of the interfaces only consisted of integers.

\subsubsection*{CNN with dimension reduction}
The architecture of the third model was designed by Liefers et al. \cite{Liefers2019} to return a one-dimensional array of $y$-coordinates for a two-dimensional image as network input. The model consists of a downsampling and upsampling path to incorporate a large contextual region. While U-Net uses direct shortcut connections to provide local context, this architecture resorts to so-called funneling subnetworks between the downsampling and upsampling path to resolve the mismatch in activation map height. The original network architecture was designed for images of 512~by~512 pixels. To avoid unnecessary computations, we adapted the architecture to work for images of 512~by~256 pixels. The downsampling path and all funneling subnetworks therefore contain one less downsampling operation and the upsampling path one less upsampling operation. $1\times1$ residual blocks at the lowest level were replaced by $1\times3$ residual blocks in our architecture. Furthermore, we experimented with the addition of batch normalization layers, which did not result in improved performance. 

As a preprocessing step, the original images were cropped twice to a size of 512~by~256 pixels for the superior and inferior half of the cornea with some overlap. Since the annotations did not span the entire width of 512 pixels, the network output layer was cropped to only include the annotated positions. 
The mean squared error between the annotation and predicted delineation was used as loss function. Training was done for 200 epochs with a learning rate set to 0.0002 at the start and divided by two after every 50 epochs. Based on preliminary experiments, we used the following data augmentation: B-scans were translated ($\leq$10 pixels) or rotated ($\leq$12 degrees) before inferior and superior crops were made. We also added uniform noise ($\leq$0.05), Gaussian blurring with $\sigma$ $\leq$1, and sigmoidal contrast changes with a gain between [4,~5].
For evaluations on the test set, the superior and inferior crops were processed and combined by averaging the central overlapping section.

\subsection*{Thickness measurements \& evaluation}
Outputs of the deep learning models were $y$-values, describing the height of the anterior and posterior interface for each $x$-coordinate within the central 9~mm. The posterior interface of the cornea generally contains more irregular shapes after DMEK. We therefore measured corneal thickness perpendicularly to the anterior interface (see Figure \ref{fig:thickness}), similar to \cite{Li2006}. First, the coefficient of proportionality was determined for the anterior interface. A 71 point moving average filter was used to reduce the effect of small deviations from the general curvature. We found that the proportionality coefficient was still affected by local irregularities after filtering with smaller filters, whereas larger filters introduced inaccuracies near the sides. The distance was then measured perpendicularly to a tangent with the corresponding coefficient of proportionality for every pixel on the anterior interface inside a 9~mm diameter. 

Performance of the models was measured by comparing the thicknesses predictions with the thicknesses following from the three sets of manual annotations. The mean absolute error was then calculated for a 3~mm, 6~mm, and 9~mm diameter. To obtain a measure of variation, we trained all models five times from scratch using different random seeds. Mean absolute errors and sample standard deviations shown in Table \ref{Tab:MAD-distance} and Table \ref{Tab:Training_size} were based on these five repetitions.

For assessment of the interface delineations errors by the built-in software of the OCT system, all B-scans of the test set were semi-quantitatively assessed by a cornea specialist. With the built-in drawing tool, missed parts of the cornea or areas mistakenly classified as cornea were selected. This was done approximately for the central 9~mm diameter although these B-scans were not centered or horizontally aligned. Sometimes the posterior delineation was not complete for the peripheral cornea. In such cases no extra misclassified area error was added; these regions were simply ignored. B-scans with a total incorrectly classified area of more than 0.1~mm$^{2}$ were considered to contain a clinically relevant inaccuracy. When this area was larger than 0.25~mm$^{2}$ the inaccuracy was considered severe.

\subsection*{Reduced training set}
Partitions of the training set were made by randomly selecting all B-scans from a subset of the study participants. For 50\% of the training set this equaled 24 participants (384 B-scans). For 25\% 12 participants (192 B-scans) and for 10\% 5 participants (80 B-scans). Partitions were randomly sampled for each of the five training repetitions, and partitions were the same for each deep learning model. All other training parameters were similar as for the 100\% training set size models.

\bibliography{main}

\section*{Acknowledgements}
This  research  is  financially  supported  by  the  TTW  Perspectief  program  and  Philips  Research.

\section*{Author contributions statement}
F.H, M.A. and M.V. designed the study. M.A. and J.C. provided the material and approved the study. R.L, M.B, L.H. and F.H. implemented the models, conducted the experiments, and analyzed the results. M.A, J.C, J.P, and M.V reviewed the analysis. All authors reviewed the manuscript.

\section*{Competing interests}
The author(s) declare no competing interests.

\section*{Additional information}
\textbf{Correspondence} and requests for materials should be addressed to F.H.
\end{document}